# Atomistic Investigation on the Mechanical Properties and Failure Behavior of Zinc Blende Cadmium Selenide (CdSe) Nanowire

Emdadul Haque Chowdhury[1], Md. Habibur Rahman[1], Md Mahbubul Islam[2]

[1]Department of Mechanical Engineering, Bangladesh University of Engineering and Technology, Dhaka-1000, Bangladesh

[2]Department of Mechanical Engineering, Wayne State University, Detroit MI – 48202, United States

*Corresponding Author. E-mail address: mahbub.islam@wayne.edu[2]

## Abstract

The mechanical properties of Cadmium Selenide (CdSe) nanowire is an emerging issue due to its application in semiconductor and optoelectronics industries. In this paper, we conducted molecular dynamics (MD) simulations to investigate the temperature-dependent mechanical properties and failure behavior of Zinc-Blende (ZB) CdSe nanowire under uniaxial tensile deformation. We employed Stillinger-Weber (SW) potential to describe the inter-atomic interactions. The effect of variation of temperatures (100 K-600 K), sizes, and crystal orientation on the tensile response of the CdSe nanowires is investigated. Our simulation results suggest that both ultimate tensile strength and Young's modulus of CdSe have an inverse relationship with temperature. From 100K to 600K, the ZB CdSe exhibits brittle type failure thus there is no brittle to ductile transition temperature found. Results also suggest that size has a significant effect on the mechanical properties of CdSe nanowire. It has been found that as the cross-sectional area increases both ultimate tensile stress and Young's modulus increases as well. The [111] oriented ZB CdSe shows the largest ultimate tensile strength, Young's modulus and fracture toughness whereas the values are lowest for [100] orientation. The [110] orientation shows the largest failure strain compared to other orientations. Finally, failure mechanisms of CdSe nanowire are also investigated at 100K and 600K. We noticed that at 100K temperature [100] oriented ZB CdSe fails along {111} cleavage plane however in the case of 600 K temperature, both {111} and {100} planes are activated and cause fracture of CdSe nanowire at lower strain value. This study can guide to design ZB CdSe based solar cell, optoelectronic and semiconductor devices by presenting a comprehensive understanding of the mechanical and fracture characteristics of this nanowire.

## 1. Introduction

Low-dimensional nanomaterials are attracting intense research interest because of their unique anisotropic and dimension-tunable electronic properties, as well as for their potential applications in the areas of electronics, optics, sensors, and semiconductor nanostructures [1, 2, 3]. The material properties of nanowires (NW) have been extensively studied for their use as building blocks in nanoelectromechanical devices (NEMS), owing to their distinctive thermal, mechanical, electronic and optical properties [4-6]. They are excellent candidates for ultrahigh-frequency resonators [7], detectors and sensors for biological and chemical applications [8], energy harvesting [9], optoelectronics and nanoelectronics devices [10].

The suitable band gaps and high absorption coefficients make Cd-based compounds such as CdTe and CdSe as the most suitable photovoltaic materials available for low-cost high-efficiency solar cells. Besides, because of their large atomic number, Cd-based compounds like CdTe and CdZnTe have been applied to radiation detectors. The attraction of Cd-based binary and ternary compounds arises from their widespread applications as solar cells, γ- and x-ray detectors, etc. These devices made from Cd-based materials are being largely applied in many fields [11]. Cadmium selenide (CdSe), an important II–VI semiconductor, has attracted a lot of attention because of its special nonlinear optical properties, luminescent properties, quantum-size effect, the bandgap in visible light range and some other superior properties [12-16].

Recently, a wide variety of CdSe nanostructures including NWs, nanocrystals and nanospheres have been synthesized, and their structural, optical, and electric properties are deeply investigated [16-18]. Among the Cd-based nanosystems, CdSe nanowires are the most appealing to researchers and scientists since they have distinctive optoelectrical properties stemming from a high aspect ratio and surface/volume ratio. The CdSe NWs are generally a mixture of Zinc-Blende (ZB) and Wurtzite (WZ) structure and the proportion of the two types of NWs depends on the synthesis conditions [19-22]. Although many efforts have been put to examine the electronic and optical properties of CdSe nanowire [23,24], few experimental and theoretical information is available for the inspection of mechanical properties of CdSe nanowire. Understanding the mechanical behavior of CdSe nanowire is very important to determine the strength of these materials for practical applications as optical or electronic interconnects and as components in nano electromechanics. Very recently, several theoretical researches have investigated the mechanical properties of WZ type CdSe nanowire [25-26], but to our best knowledge, the studies on ZB CdSe nanowire are very scarce. The effect of diameter and temperature on the tensile mechanical properties of ZB CdSe nanowire has been studied by Fu et al [27]. They found that as the temperature and diameter of the nanowire increases, the ultimate tensile strength and Young's modulus decrease. However, a detailed understanding of crystallographic direction dependent mechanical properties, and failure mechanisms of the ZB CdSe NW at different temperatures are remain elusive.

Molecular dynamics (MD) simulation is a powerful tool to analyze the mechanical behavior and deformation mechanisms of nanomaterials under different loading conditions and this method has been largely used and proved to be an effective tool in the inspection of the nanomechanical behavior of various nanowires [25-28]. In this study, we used MD simulations to study the size, temperature, and crystal orientations dependence on the tensile properties and failure mechanisms of ZB CdSe nanowire. We elucidate the effect of temperature, size and crystal orientation on ultimate tensile strength and Young's modulus of ZB CdSe nanowire. Finally, the fracture mechanism of ZB CdSe at extremely low and high temperatures are investigated as well. We envisage that this work can detail a pathway in the understandings of mechanical characteristics of CdSe nanowire, which may be useful for experimental work and understanding other properties of CdSe nanomaterials.

## 2. Computational Method

We used molecular dynamics (MD) simulations using the Large-scale Atomic/Molecular Massively Parallel Simulator (LAMMPS) package for calculating the tensile properties of CdSe nanowire [36]. To describe the interatomic interactions between Cd-Se systems, we employed Stillinger-Weber (SW) potential [39]. The SW potential consists of a two body term and a three-body term describing the bond stretching and bond breaking interactions respectively [31]. The mathematical expressions of these interactions are described below, [31]

$$\varphi = \sum_{i<j} V_2 + \sum_{i<j<k} V_3$$

$$V_2 = \varepsilon A(B\sigma_a^p r_{ij}^{-p} - \sigma_a^q r_{ij}^{-q})e^{[\sigma_a(r_{ij}-a_1\sigma_a)^{-1}]}$$

$$V_3 = \varepsilon A e^{[\gamma\sigma_a(r_{ij}-a_1\sigma_a)^{-1}+\gamma\sigma_a(r_{jk}-a_1\sigma_a)^{-1}]} \cdot (\cos\theta_{ijk} - \cos\theta_0)^2$$

Here, two-body and three-body terms are denoted by $V_2$ and $V_3$ respectively; $r_{ij}$ denotes the distance between atoms $i$ and $j$; $\theta_{ijk}$ represents the angle between bond $ij$ and $jk$; equilibrium angle between two bonds is denoted by $\theta_0$; $A$, $B$, and other parameters are the coefficients required to fit while developing the interatomic potential function. The values of these parameters can be found in reference [40].

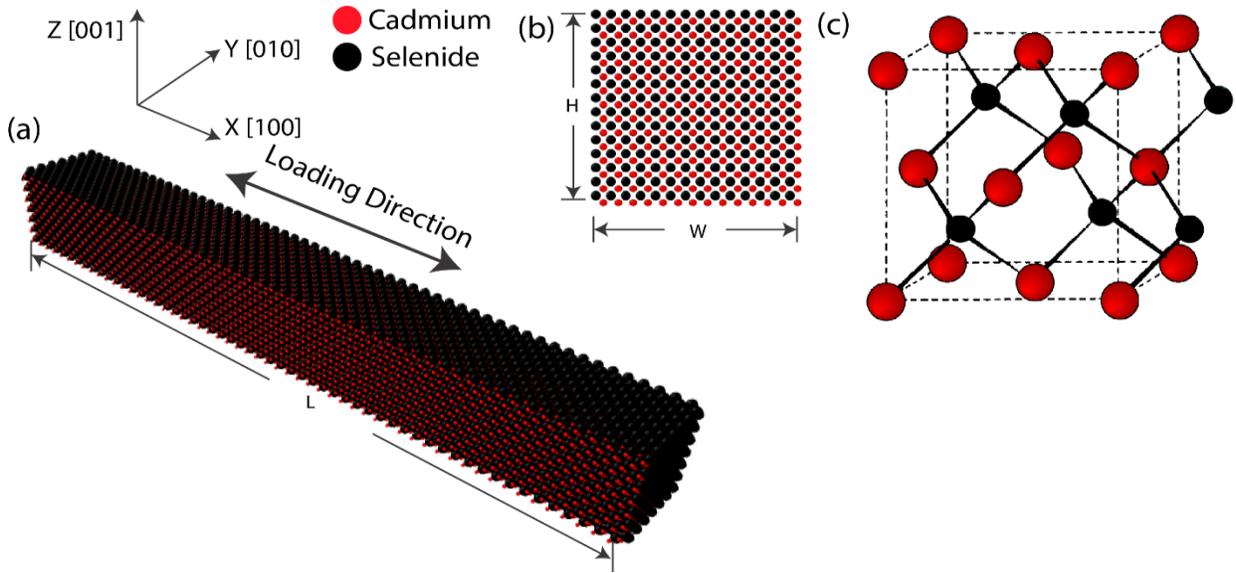

**FIGURE 1**. (a) Initial nanostructure of [100] ordinated Zinc Blende (ZB) 33.936 nm x 4.242 nm x 4.242 nm CdSe nanowire (c) Cross-section of nanowire (b) ZB crystal structure of CdSe

We investigated the effect of variation in temperature and cross-sectional area. We considered a temperature range of 100K to 600K and square cross-sectional areas are varied from 13.22 nm² to 23.50 nm² with a constant length to width ratio of 8:1 [31,32]. Three different crystal orientations such as [100], [110] and [111] are chosen for this investigation [32]. At first, we performed structural relaxation using Conjugate Gradient (CG) minimization scheme [31] followed by constant energy (NVE) simulations for 40 ps. Next, NPT simulations are carried out at atmospheric pressure for 40 ps to eliminate any residual stresses in the structure with a temperature and pressure damping parameters of 0.5 ps and 0.5 ps, respectively. Then, canonical NVT ensemble is performed for 10 ps [32]. Finally, uniaxial strain is applied along X direction at a constant strain rate of $10^9\ s^{-1}$ under NVT ensembles to control temperature fluctuations [31,32]. Periodic boundary condition is applied along the loading direction and other directions are kept nonperiodic [31]. This strain rate is higher than the adopted in real life because we are constrained by the computational resources. But previous works have found that the strain rate of $10^9\ s^{-1}$ is good enough for atomistic simulations [31, 32]. The atomic stress in our simulation is calculated using Virial stress theorem [31,32] which stands as,

$$\sigma_{virial}(r) = \frac{1}{\Omega}\sum_i(-\dot{m}_i\dot{u}_i \otimes \dot{u}_i + \frac{1}{2}\sum_{j\neq i}r_{ij} \otimes f_{ij})$$

In this theorem, the summation is performed over all the atoms occupying a volume, $\dot{m}_i$ presents the mass of atom i, $\otimes$ is the cross product, $\dot{u}_i$ is the time derivative, which indicates the displacement of the atom with respect to a reference position, $r_{ij}$ represents the position vector of the atom, and $f_{ij}$ is the interatomic force applied on atom i by atom j [31,32]. OVITO is used for all the visualization [37].

To validate the interatomic potential used in this study, we calculated Young's modulus of ZB CdSe nanowire at 300 K and compared them with published literature [13].

Table 1: Comparisons of mechanical properties between present calculation and existing literature

| Property | This study | Literature |
|---|---|---|
| Young's modulus of CdSe NW | 58 GPa | 62 GPa [13] |
| Lattice constant of CdSe | 6.03 $\mathring{A}$[13] | 6.05 $\mathring{A}$[13] 6.05 $\mathring{A}$[38] |

The results demonstrate the capability of the potential to accurately predict the mechanical properties of the CdSe nanostructures.

## 3. Results and Discussions

**3.1 Effect of temperature and cross-sectional area on the mechanical properties of ZB CdSe:** The stress–strain response for 18nm$^2$ ZB CdSe nanowire under uniaxial tensile loading for 100 K-600 K is shown in Fig. 2(a). From stress-strain diagram it is evident that up to certain point (3% strain) stress increases linearly with the increasing strain thus material exhibits elastic behavior in that region. After that, stress increases non-linearly and reaches its peak called ultimate tensile strength (UTS) and then decreases abruptly with a small increase in strain [31,32]. The sharp decrease in stress indicates a brittle-type failure of the CdSe. Moreover, at 100 K, the ultimate strength is calculated as 8.55 GPa, with a failure strain of about 18.75%. As the temperature increased from 100 K to 600 K, the ultimate tensile strength of CdSe decreased to 6.66 GPa, with a lower value of failure strain of about 16.25%. This behavior can be attributed to the temperature-induced weakening of the chemical bonds in the crystal and higher thermal vibrations at elevated temperatures.

To observe the effects of size on the tensile behavior of ZB CdSe nanowire, the cross-sectional area of the nanowire was varied from 13.22 nm$^2$ to 23.50 nm$^2$. The length of the nanowire was varied from 29.08 nm to 38.78 nm maintaining a length to width ratio of 8:1. The results of three simulations for different cross-sectional area at 300 K are plotted in Fig. 2(b). It is evident from this figure that the stress-strain response depends on the cross-sectional area of CdSe nanowire. These results are translated into Fig. 2(c-d), which represents the variations of ultimate tensile strength and Young's modulus with temperature for different cross-sectional area. Young's Modulus is obtained by fitting the stress-strain curve to a straight line and strain value less than 3%. One can also observe from these two figures that there is significant impact of size on the mechanical properties of [100] oriented CdSe nanowire. From Fig. 2(c-d), it is evident that both ultimate tensile strength and Young's modulus show an inverse relation with temperature for all cross-sectional areas. It is also evident that as cross-sectional area increases both ultimate tensile strength and Young's modulus of CdSe increases as well. However, the effect of size is more prominent at higher temperatures. It is generally observed that surface to volume ratio is mainly accountable for the size dependent behavior of nanowires failure [33]. Therefore, we can say that the surface atoms have a significant effect on the tensile properties of CdSe nanowire. In addition, the smaller cross-sectional nanowires have comparatively larger specific surface area, and the lattice defects mainly occur on the surface of the nanowires. Therefore, nanowires having smaller cross-sectional areas have higher ratio of lattice defects compared to larger cross-sectional areas [33]. The degree of lattice defects of the nanowires directly affects the ultimate tensile stress and Young's modulus of nanowire.

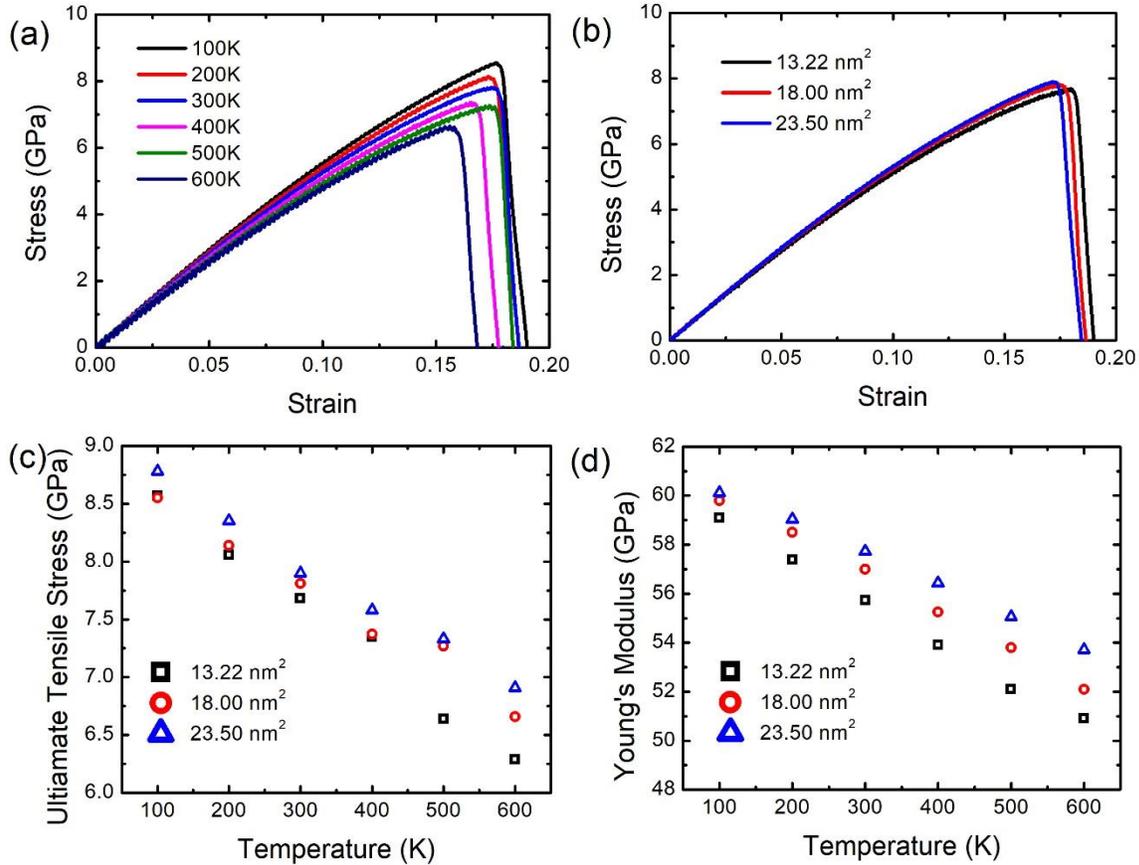

**FIGURE 2.** (a) Stress-strain curve of 18 nm$^2$ ZB CdSe for different temperatures (b) Stress-strain curve of ZB CdSe for different cross-sectional area at 300K. (c) Variations of ultimate tensile stress with temperature for different cross-section. (d) Variations of Young's Modulus with temperature for different cross-sectional area.

The 13.22 nm$^2$ nanowire has the lowest Young's modulus and ultimate tensile stress. If the cross-sectional area is further reduced, the nanowire structure become unsteady [33]. This kind of size dependent mechanical properties is also reported for CdTe nanowire [31]. From 100K to 400K the deviation between of ultimate tensile strength among three cross-sections is low but in case of 500K to 600K the standard deviation is high which is also evident in the case of Young's Modulus of ZB CdSe. For example, at 300K the standard deviation of ultimate tensile strength among the three different cross-sectional area is calculated as 0.09 GPa but in the case of 600K it is 0.25 GPa. The lower fracture toughness and Young's modulus of CdSe nanowire at higher temperatures may be due to the fact that the length of the chemical bonds between Cd-Se fluctuate with increasing temperature from their mean position [31,32]. This increases the possibility that bonds will reach the critical bond length condition, which leads to fracture of CdSe [31,32].

**3.2 Effect of crystal orientation on the mechanical properties of ZB CdSe for different temperature:**

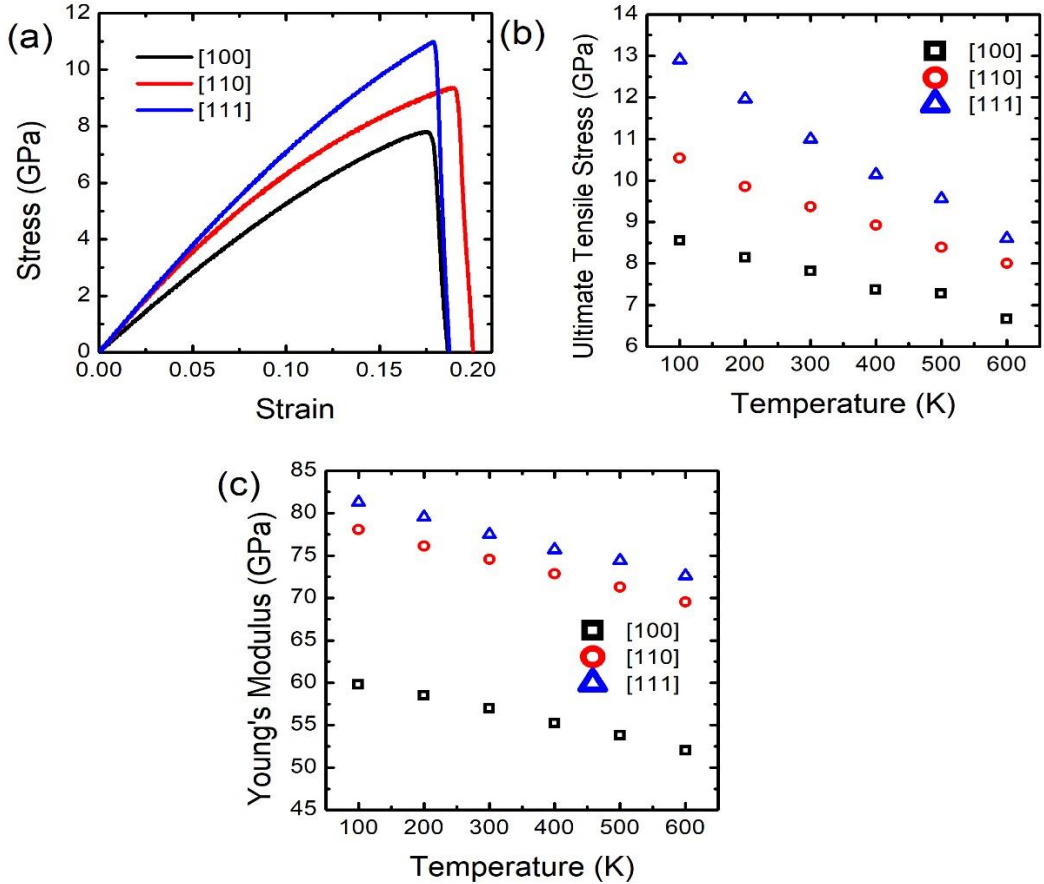

**FIGURE 3.** (a) Stress-strain curve of 18 nm$^2$ ZB CdSe for different crystal orientations at fixed 300K (b) Variations of ultimate tensile stress with temperature for different crystal orientations. (c) Variations of Young's Modulus with temperature for different crystal orientations.

Figure 3(a) shows the stress-strain curves of ZB CdSe at 300K temperature with [100], [110] and [111] crystal orientations. It can be seen that [111] oriented ZB CdSe shows largest ultimate tensile stress and the lowest one is seen for [100] orientation. This kind of orientation dependent mechanical properties are also reported for Si and CdTe nanowires [31,35]. The [110] oriented ZB CdSe shows higher fracture strain compared to other two crystal orientations. Area under the stress-strain curve calculations show that, it is the largest for [111] orientation. Therefore, the total energy absorbed before failure, that is, the fracture toughness is the maximum for [111] orientation and the minimum for [100] orientation. Figures 3(b-c) represent variations of ultimate tensile stress and Young's Modulus for different crystal orientation. The Young's modulus of [100], [110] and [111] oriented CdSe are obtained by fitting the stress-strain curve to a straight line and strain value less than 3% as mentioned earlier [31,32]. From Fig. 3(b) it is evident that the rate of decrement of ultimate tensile stress with temperature is the largest for [111] orientation and for [100] orientation it is the lowest. In the case of Young's modulus, decrement rate with temperature is nearly constant for three crystal orientations.

In the [111] oriented ZB CdSe, two types of polarities have been found in the atomic layers. From Figure 6(c), it is evident that all atoms in a layer are positively charged ($Cd^{2+}$) and all atoms in the adjacent layer are negatively charged ($Se^{2-}$). As a result, there induces a strong electrostatic attraction between two adjacent planes which makes it difficult to separate when tensile load is applied along the [111] orientation [31].

### 3.3 Failure mechanism of [100] oriented ZB CdSe at 100 K and 600 K

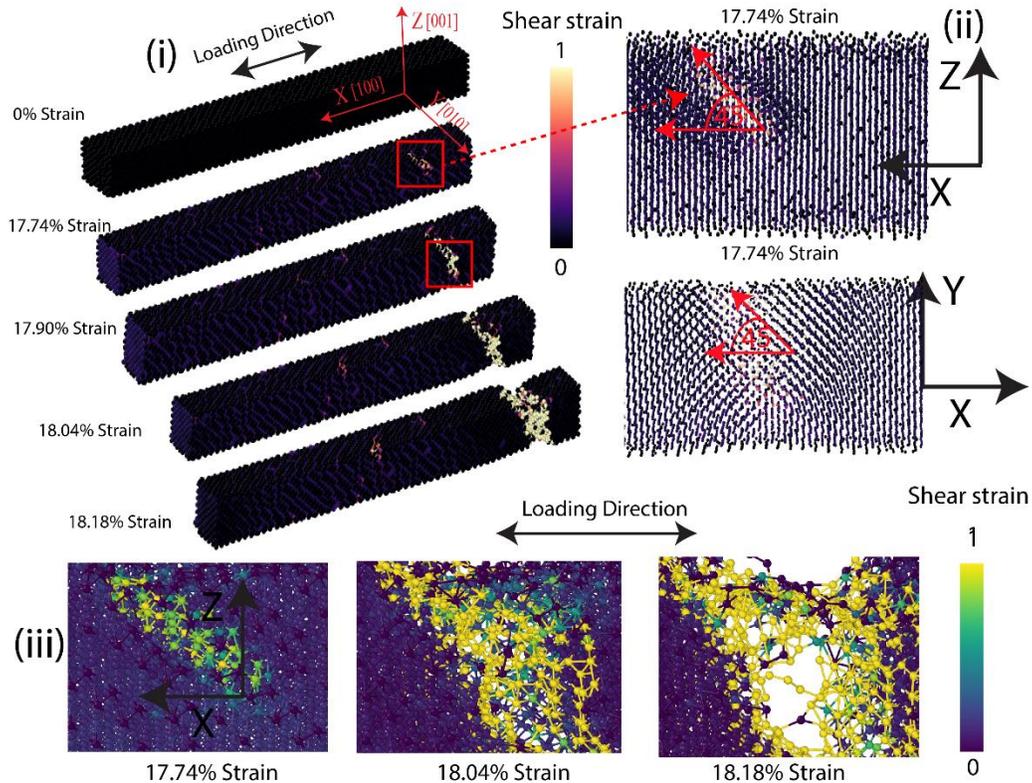

**FIGURE 4.** (i) Failure mechanism of [100] oriented 18 $nm^2$ ZB CdSe nanowire for various strain level at fixed 100K (ii) Crack nucleation from different perspective at 17.90% strain, atoms size is reduced for better visualization with the aid of OVITO (iii) Bond breaking between Cd-Se under tensile loading for various strain level. Yellow color indicates bonds that are about to break under tensile loading

The ZB CdSe nanowire in the present study failed by cleavage fracture on the {111} plane at temperature 100K. In Fig. 4, the failure of a nanowire with a cross-sectional area of 18 $nm^2$ is shown at a temperature of 100 K with the aid of the shear strain parameter. Lower values of shear strain represent a perfect ZB lattice, while higher values indicate a local defect or surface atoms. At 17.74% strain, a crack is initiated at the nanowire surface, which is shown in Fig. 4(i). It is found that fracture occurs in a strain region of 17.74% to 18.18%. To understand the failure mechanism profoundly, atomic arrangement analysis of [100]-oriented ZB CdSe nanowire is used which is shown in Fig. 4(ii). It is observed that the crack starts to branch immediately after the crack nucleation and primary displacement of atoms occur along the x axis.

After that, the crack propagates along a plane making $45^0$ angle with the nanowire axis. The $45^0$ angles of the cleavage planes with nanowire axes both from x-y and x-z planes represent cleavage fracture along the {111} plane. This kind of {111} cleavage fracture is reported earlier for InP and SiC [32,34]. For 600K at 16.15% strain, cracks started to initiate at two different places of the nanowire, which is shown in Fig. 5(i).

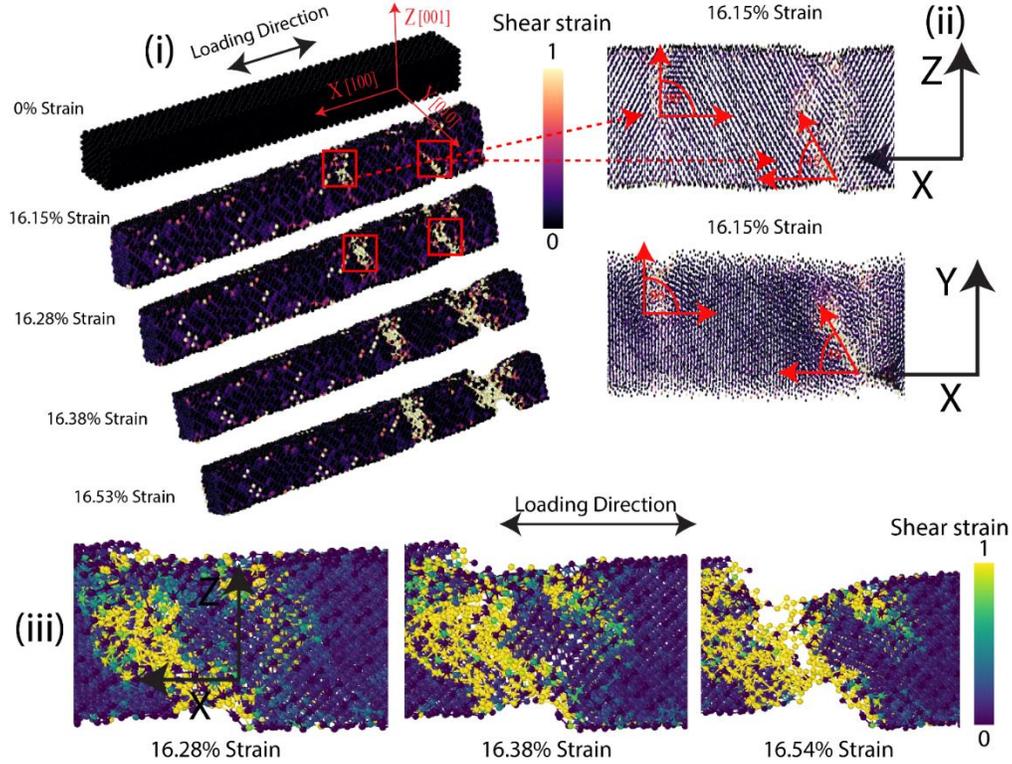

**FIGURE 5.** (i) Failure mechanism of [100] oriented 18 nm$^2$ ZB CdSe nanowire for various strain level at fixed 600K (ii) Crack nucleation from different perspective at 17.90% strain, atoms size is reduced for better visualization with the aid of OVITO (iii) Bond breaking between Cd-Se under tensile loading for various strain level. Yellow color indicates bonds that are about to break under tensile loading.

At about 16.28% strain bonds started to break and at 16.54% strain, the nanowire completely fails as the bonds between Cd-Se are almost broken. The point to be noted here that, at 600K temperature, crack nucleation, propagation and eventually the failure occurs at a lower strain values compared to 100K. It is because at higher temperature the chemical bonds in the ZB CdSe crystal experience higher thermal fluctuation and vibration [31,32]. To illustrate the cleavage planes, atomic displacement analysis is shown in Fig 5(ii). At 16.15% strain, a crack is nucleated at two different locations, which are shown with red blocks in Fig. 5(i). Upon further stretching, the cracks started to propagate at two different directions which are shown in Fig. 5(ii). From the atomic displacement analysis of Fig 5(ii), it is clear that, one of the crack planes makes $45^0$ angles in the x-z and y-z planes. It represents cleavage failure along {111} plane. The other one starts to propagate along a plane perpendicular to x direction viewing from both of the planes and it represents cleavage failure at {100} plane. Bond breaking in the cleavage plane primarily depends on surface polarity, atomic coordination number, and bond length [30]. Plane spacing and electrostatic forces are the two key factors to regulate the cleavage plane behavior of ZB CdSe nanowire. It can be seen from Fig. 6 (b-c) that, the atoms in the {110} and {111} planes are singly bonded to the opposite surface or atomic layer. The plane spacing of the {110} planes is shorter than the spacing of the {111} planes, as shown in Fig. 6(b-c). Therefore, less energy is required to nucleate a cleavage along the {111} planes because they have larger spacing than the {110} planes. In contrast, in the {111} planes, there are two types of polarities in the atomic layers. From Fig. 6(c), it can be observed that all atoms in a layer are positively charged ($Cd^{2+}$) and all atoms in the adjacent layer are negatively charged ($Se^{2-}$).

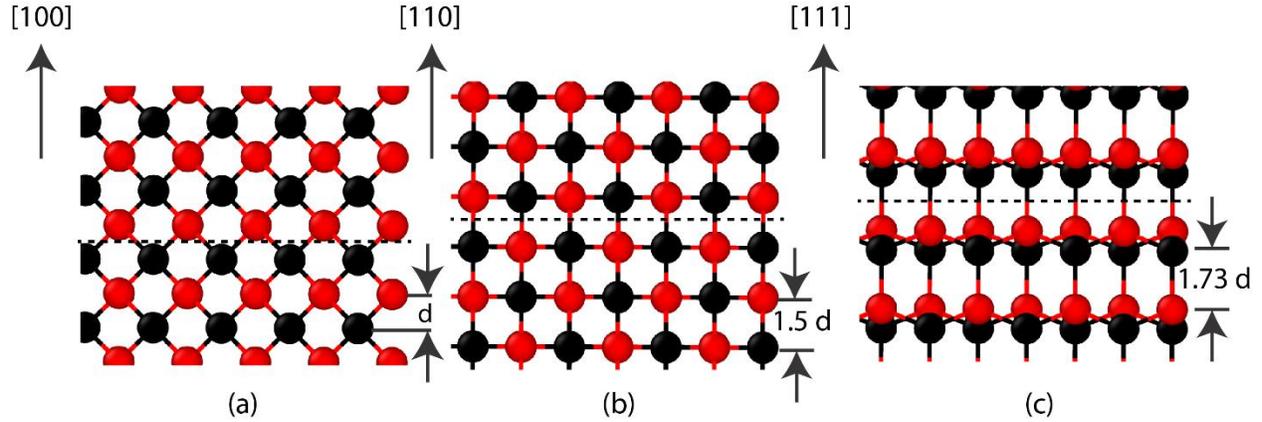

**FIGURE 6**. Representation of the atomic arrangements of ZB CdSe in the direction along the (a) [100], (b) [110], and (c) [111]. The relative spacing's between the $Cd^{2+}$ and $Se^{2-}$ atoms is also represented. Red color represents Cd atom and black color represents Se atom.

For that reason, there will be a strong electrostatic attraction between two adjacent planes due to oppositely charged atoms ($Cd^{2+}$ and $Se^{2-}$), for which it is difficult to separate the bonds along the {111} plane. However, the {110} layers are composed of equal numbers of $Cd^{2+}$ and $Se^{2-}$ atoms, resulting in poor electrostatic forces. Interestingly, in the case of CdSe nanowire the plane spacing effect is dominant over the electrostatic forces and hence, both at 100K and 600K temperatures, cracks propagate along {111} plane. At, 600K, cleavage plane {100} also becomes active. Along {100} plane, the bonds are arranged in such a way that all the bonds make 45˚ with the loading direction (indicated by the arrow in Fig. 6(a)). As a result, these bonds fail at a lower tensile and compressive load as reported in the previous literature [31]. They have found that, in [100] direction the maximum value of stress and the Young's modulus is lower compared to [111] and [110] direction. Though breaking bonds along the {100} plane should require enormous energy because of its shorter bond length. Additionally, every atom must break two bonds to nucleate a crack (Fig. 6(a)). When the temperature is low, the structure of the nanowire shows little or no movement from its equilibrium position. Therefore, the shorter bond of the {100} planes dominantly affects the failure. That is why breaking bonds in this plane at lower temperature is not observed. However, at elevated temperature, the amplitudes of atomic vibration become higher and atoms can move from their equilibrium position more easily. For this reason, the effect of atomic spacing loses its prominence at higher temperature [31,32]. As a result, at 600K along with {111} plane, the nanowire fails at {100} plane as well.

## 4. Conclusions

In summary, we investigated the mechanical properties and failure mechanism of ZB CdSe nanowire at the atomistic level by using molecular dynamics simulation. The effects of variation of temperature (100K-600K), nanowire size, and different crystal orientation on the tensile behavior of this nanowire are comprehensively analyzed. It is demonstrated that, the temperature has a significant effect on the stress-strain behavior of CdSe NW. For all different temperatures, the nanowire shows a typical brittle type behavior and no ductile to brittle transition temperature is not found. We also found that both the ultimate strength and Young's modulus of the NW show an inverse relationship with temperature. The effect of the NW size is found to have a negligible effect on the stress-strain behavior of CdSe nanowire below room temperature but the effect is prominent at higher temperatures. One of the major findings of this study is that, the mechanical properties of ZB CdSe are strong function of crystallographic orientations. In the entire temperature range, [111] direction shows the maximum ultimate strength, Young's modulus and fracture toughness whereas the lowest values are obtained for [100] direction. However, the maximum fracture strain is obtained for [110] direction. Our study also reveals the temperature dependent failure behavior of ZB CdSe NW. At low temperatures, only {111} cleavage planes are observed. However, at elevated temperature, surprisingly, along with {111} plane, {100} plane also becomes activated. It is also observed that this temperature-dependent cleavage plane behavior of ZB CdSe nanowires is controlled by competition between atomic spacing and electrostatic forces between the cleavage planes. This investigation details an extensive understanding of temperature, size, and crystallographic orientation-dependent mechanical behavior and fracture phenomenon of ZB CdSe NW which has enumerable application in optoelectronic and semiconductor industries.

# Acknowledgement

Authors of this article would like to thank Department of Mechanical Engineering, Bangladesh University of Engineering and Technology (BUET) and Multiscale Mechanical Modeling and Research Networks (MMMRN) for providing supports. MMI acknowledges start-up funds from Wayne State University. We sincerely acknowledge the contribution of Mr. Pritom Bose for his insightful comments on the paper.